\documentclass{aastex631}
\usepackage{amsmath}

\begin{document}

\title{The First Model-Independent Chromatic Microlensing Search: No Evidence in the  Gravitational Wave Catalog of LIGO-Virgo-KAGRA}

\author[0009-0004-4937-4633]{Aniruddha Chakraborty}
\affiliation{Tata Institute of Fundamental Research, 
Homi Bhabha Road, Navy Nagar, Colaba, 
Mumbai 400005, India}
\email{aniruddha.chakraborty@tifr.res.in}

\author[0000-0002-3373-5236]{Suvodip Mukherjee}
\affiliation{Tata Institute of Fundamental Research, 
Homi Bhabha Road, Navy Nagar, Colaba, 
Mumbai 400005, India}
\email{suvodip@tifr.res.in}



\begin{abstract}
The lensing of Gravitational Waves (GWs) due to intervening matter distribution in the universe can lead to chromatic and achromatic signatures in the wave-optics and geometrical-optics limit respectively. This makes it difficult to model for the unknown mass distribution of the lens and hence requires a model-independent lensing detection technique from GW data. We perform the first model-independent microlensing search in the wave-optics limit on the 72 super-threshold GW events observed with both the LIGO detectors up to the third observation catalog GWTC-3 of LIGO-Virgo-KAGRA using the analysis method $\mu$-\texttt{GLANCE}. These unmodelled searches pick up one plausible candidate $\rm GW190408\_181802$ with a slightly above threshold residual amplitude compared to the residual expected from detector noise. However, exploring the microlensing modulation signatures on this event, we do not find any conclusive evidence of the microlensing signal in the data. With this, we confidently rule out the presence of any statistically significant microlensing signal in the 72 events up to GWTC-3 in a model-independent way.

\end{abstract}


\section{Introduction} \label{sec:intro}
Gravitational waves (GWs) are propagating spacetime perturbation, emanated from a time-varying matter distribution. When GWs traverse by a massive astronomical object they get deflected, time-lagged and amplified. This is known as the lensing of gravitational waves. If the lensing astronomical object is of similar size \footnote{The `size' refers to the Schwarzschild radius of the lens, $R_s = \frac{2GM_{z, lens}}{c^2}$, where $M_{z, lens}$ is the redshifted mass of the lens and $G$, $c$ be usual constants. Here $M_{\rm z, lens}$ is the redshifted mass of the lens given by, $M_{\rm z, lens} = M_{\rm lens} (1+z_{\rm lens})$ with $z_{\rm lens}$ being the redshift of the lens. G and c denote usual constants.} or comparable to the wavelength of the GW, we can observe diffraction-effects due to the pronounced wave nature of the GWs. This range sets the size of the lenses which can be probed with wave-optics lensing of GWs. The regime is known as the microlensing (wave-optics lensing) of gravitational waves \citep{1992grlebookS, Takahashi:2003ix}. Typical microlensing effects on the GW include frequency-dependent amplitude and phase modulation, although the frequency evolution, which is dependent of the chirp mass of the compact binary, is unaltered by lensing \citep{Diego:2019lcd, 2021MNRAS.503.3326C}. However, in-plane spin components of the black holes is affected by the presence of microlensing \citep{Mishra:2023ddt}. Amplification effects in the wave-optics regime, especially with strong frequency-dependent phase distortions, can hinder detection. Matched-filtering pipelines using unlensed templates may struggle to identify the microlensed signal \citep{Chan:2024qmb}. The features of the microlensing effect on the GW is subjective to the gravitational potential of the lensing body. Thus gravitational microlensing can offer a deep understanding about the mass distribution of the lensing object. EM wavelengths being too small than any generic astrophysical body (except radio waves), they are most often lensed in the geometric-optics way, which is unable to provide any wave-optics features coming of the lens structure. GWs, on the other hand, being spread over large range of wavelengths, can probe astronomical objects of typical size of $10^{2}$ km (in the Laser Interferometer Gravitational-wave Observatory (LIGO) band $10^3$ Hz $\equiv 10^2$ km) $-10^{14}$ km (in the Pulsar Timing Array (PTA) band $10^{-9}$ Hz $\equiv 10^{14}$ km) in the wave-optics regime. This can allow us to characterize unknown astronomical objects, such as to understand the properties of dark matter with multiband observations with GWs \citep{Massey:2010hh, Basak:2021ten, Cao:2022mrc, Fairbairn:2022xln, Tambalo:2022wlm}.

Until now, the searches of microlensing signal from GW data is performed for simple spherically symmetric lens models and without including both strong and microlensing features. From the first three observational runs (O1-O3) of the LVK detectors, there is no strong evidence of any lensing events, except a few potential candidates \citep{Hannuksela:2019kle, LIGOScientific:2023bwz}. Follow-up studies performed on some of the GW events found no conclusive microlensing candidate \citep{Dai:2020tpj, Janquart:2023mvf}. 

In this work, we applied a completely model-free approach based on residual cross-correlation search $\mu$-\texttt{GLANCE} \citep{Chakraborty:2024mbr} to search for microlensing signatures on these GW events. The method looks for any correlated feature present in the GW residuals combining different GW detectors in a completely model-independent data-driven approach. Cross-correlation between residuals is able to find any common features from microlensing while suppressing the uncorrelated noise in the detectors. The method is also extendable to strong lensing searches by cross-correlating between the data pieces themselves. This is shown in our previous work \texttt{GLANCE} \citep{Chakraborty:2024net}. We applied this wave-optics lensing search method on the GW public data. The public data of the LIGO-Virgo-KAGRA (LVK) detector network contains a total of 90 GW events \citep{KAGRA:2021vkt}, all of them are believed to come from the coalescence of compact binaries. Out of these 90 events, 72 GW events are observed with both the LIGO detectors: LIGO-Livingston and LIGO-Hanford, with a network matched-filter SNR above 8 \footnote{Matched-filter signal-to-noise ratio (SNR) estimates the noise-weighted overlap between the GW waveform template and the strain data. It is given by, $\rho_{i} = \sqrt{4\, \textbf{Re}\left[\int_{f_{\min}}^{f_{\max}} \frac{s(f) h^{*}(f)}{S_N(f)}\, df\right]}$, where $s(f)$ is the frequency-domain data and $h^{*}(f)$ is the complex conjugate of the template waveform. $S_N(f)$ denotes the noise power spectral density (PSD). The overlap is calculated in the frequency domain between $f_{\min}$ and $f_{\max}$, $\textbf{Re}$ denotes taking the real part of the complex integral. The network matched-filter SNR calculates the combination of these SNRs by adding them in quadrature. It is given by, $\rho_{N} = \sqrt{\sum_i \rho_i^2}$ ; a higher matched-filter SNR signifies higher confidence in the GW detection.}. We consider these sources for the wave-optics lensing searches using $\mu$-\texttt{GLANCE}.

The paper is organized as follows, in section \ref{sec2}, we briefly describe the basics of gravitational microlensing of GWs and the mathematical technique involved in $\mu$-\texttt{GLANCE}, to search for microlensing. It is followed by section \ref{sec3}, where we apply the technique on the publicly available GW data, we classify events into interesting and non-interesting events for microlensing. To determine the degree at which the microlensing modulations affect the signal, we apply a lens characterization through a Bayesian framework. We also emphasize the effects of waveform systematics in such analysis and how that effect is minimized in this work. In \ref{sec1} we compare our analysis with respect to the previous works and in \ref{sec4}, we state the future prospects of a microlensing detection with the inclusion of next-gen GW detectors along with the improvement of the sensitivities of the current detectors.

\section{Basics of microlensed search technique using $\mu$-\texttt{GLANCE}}\label{sec2}

\begin{figure}
    \centering
    \includegraphics[width=0.9\linewidth]{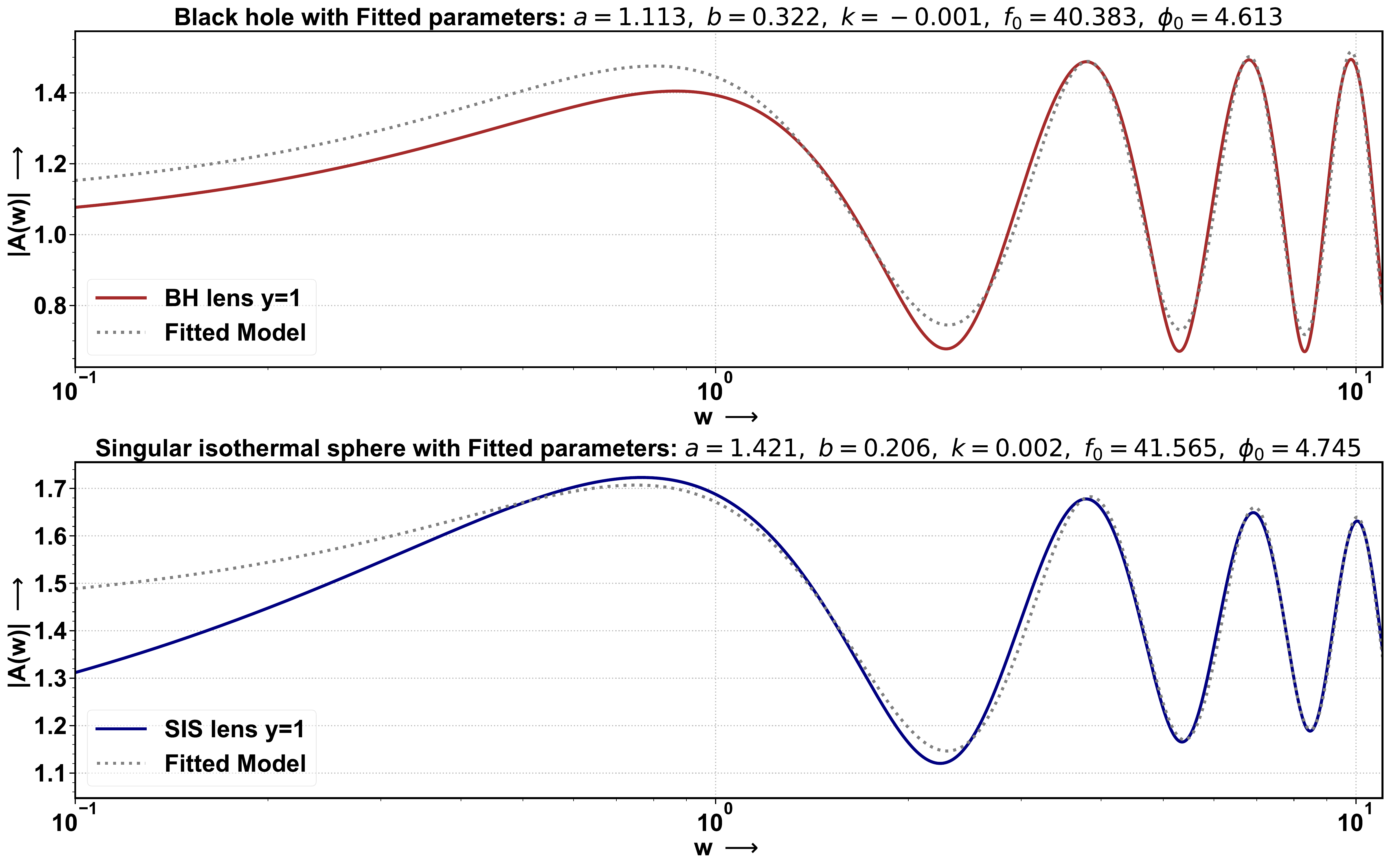}
    \caption{The figure shows the dimensionless frequency $w \equiv \frac{8 \pi M_{z, lens} f}{c^3}$ variation of the amplification  for a black hole lens system and for a singular isothermal sphere lens system in the microlensing regime. The dimensionless misalignment parameter $y$ is chosen to be unity. We observe that the toy model chosen for the lensing amplification fits quite well with both of these two models. The fitting parameters are mentioned at the top of the each panel.}
    \label{fig:amp_fit}
\end{figure}

\begin{figure}
    \centering
    \includegraphics[width=0.85\linewidth]{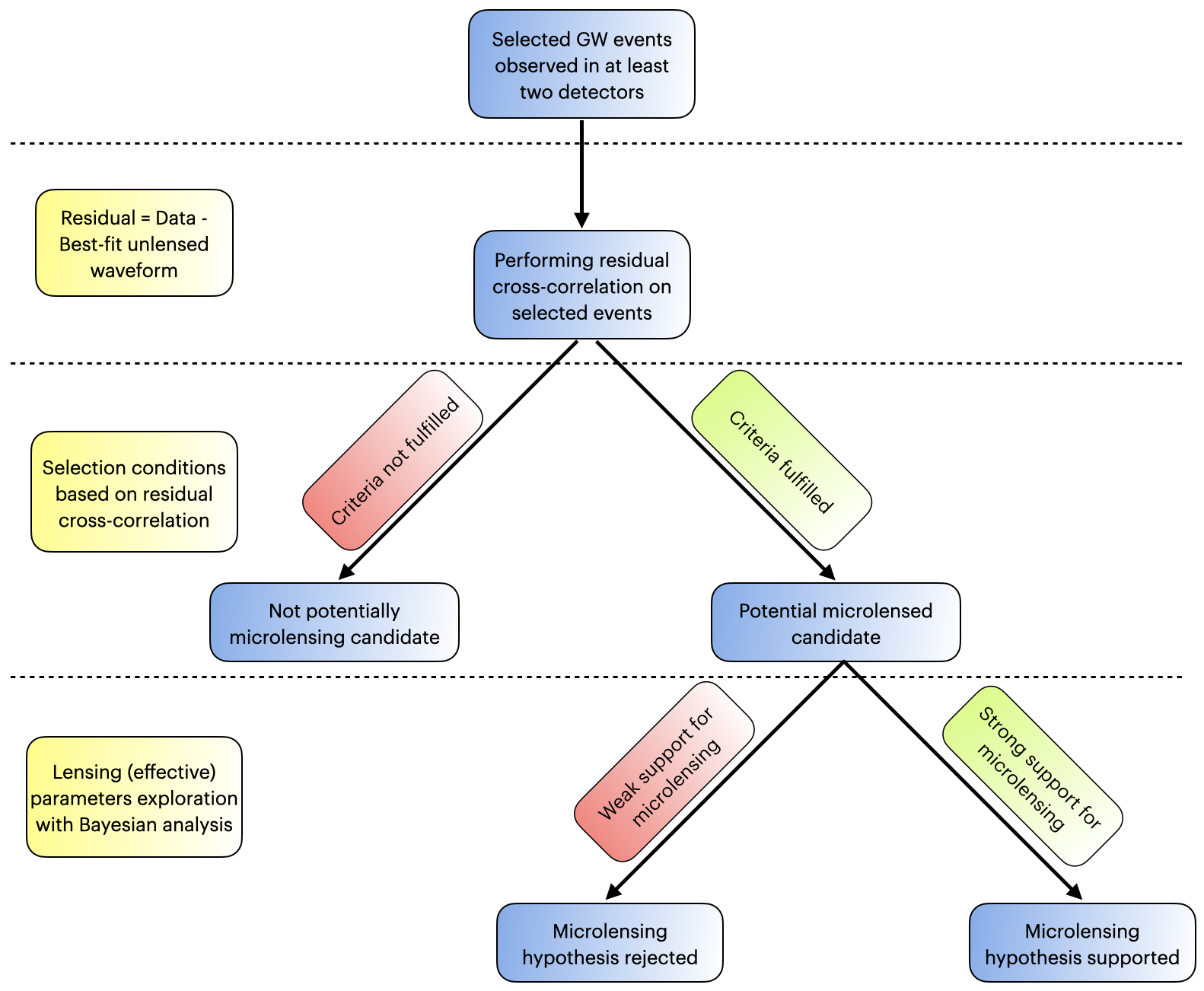}
     \caption{The figure shows the workflow for the application of $\mu$-\texttt{GLANCE} on data to search for microlensing. We select the events observed with both LIGO-Hanford (H1) and LIGO-Livingston(L1) detectors. We estimated the best-fit signal strain corresponding to the maximum likelihood parameters. The best-fit strain is used to calculate the residuals. We cross-correlate the residuals from two different detectors. We consider an event as interesting for microlensing if the cross-correlation passes through the listed selection criteria mentioned in the next section. To confirm the presence of microlensing features on the signal, we perform a Bayesian analysis to estimate the strength of the microlensing imprints on the data. We assign microlensing candidature to an event if it shows strong support for the lensing parameters.}
    \label{fig:outline}
\end{figure}

Gravitational lensing is the bending of EM waves or GWs around an object that distorts the local spacetime. These bending tend to converge the incoming wave-vectors, making them interfere. 
Gravitational microlensing in the wave-optics regime \citep{Takahashi:2003ix} occurs when the Schwarzschild radius of the lens $R_s = \frac{2GM_{\rm z, lens}}{c^2}$ \citep{Schwarzschild:1916uq} is of the order of or comparable to the wavelengths of the gravitational wave. This is the wave regime of the GWs and any lensing effect on the GW is well understood in terms of the wave-interference effects. These microlensing effects vary with the gravitational potential, which in turn varies with the mass distribution of the lens. Any generic microlensing produces frequency-dependent amplitude and phase modulation of the incoming GW. The amplification factor $A(f) = \frac{h^{\rm lensed}(f)}{h^{\rm unlensed}(f)}$ is a complex, slowly decaying function of frequency, with both phase and amplitude values oscillatory in nature. The function does not decay to zero as the frequency continues to increase; rather it converges to a positive value denoted by the strong-lensing magnification $\sqrt{\mu}$, $\mu$ denoting the flux magnification in lensing. This is known as the strong lensing regime, where both the amplitude and phase of the complex amplification factor converge to a single value \footnote{On the other end, when wavelengths are much larger than the size of the lens, we are in the diffraction-limited regime, where the amplification converges to unity. In this scenario, the wave effectively ignores the potential of the lens.}. Thus, the strong lensing effects are independent of frequency, and it only produces constant magnifications and phase-shifts. In the microlensing regime, the oscillation amplitude, periodicity, and decay rate depend on the impact parameter (how misaligned the source is compared to the straight line joining the source and the lens) and the Schwarzschild radius of the lens as compared to the wavelength of the GW \footnote{For a more detailed discussion on how micro-lens gravitational potential and the impact parameter decides the amplification, you may refer to \citep{1992grlebookS}.}. An amplification model that successfully fits all the above constraints is given in \ref{eq:lens},

\begin{equation}\label{eq:lens}
A(f) = a\left[1+b e^{-kf} \cos\left(\frac{2 \pi f}{f_0} + \phi_0 \right)\right] \hspace{0.25cm},
\end{equation}
here, the term $a$ is equivalent to the strong lensing magnification $\sqrt{\mu}$. So when the frequency $f$ becomes much larger than $1/k$, we observe no microlensing effects in the amplification. $b$ is the microlensing amplitude, with the frequency-dependent amplitude oscillates. The oscillation periodicity is set by $f_0$. The amplitude $b$ depends on the misalignment between the GW and the lensing objects and the oscillation periodicity is set by both the mass distribution of the lens and the misalignment. $\phi_0$ is the initial phase of the oscillation. The figure \ref{fig:amp_fit} shows how the modeled amplification factor matches with the two very simple lens amplification models for the black hole lens (point-mass) and for the singular isothermal sphere lens (whose density goes as $1/r^2$ from the center). We also mention the fitting values at the top of the plot.

To detect microlensing, we implement a residual cross-correlation based search technique $\mu$-\texttt{GLANCE} on the LVK public data. The method is completely unmodelled and, therefore, is able to capture the microlensing effects from astronomical objects for any mass distribution of the lens system. Given that the microlensing signature is present in the data and common to the data of all detectors, $\mu$-\texttt{GLANCE} finds any common feature present in the residuals from different detectors when the different detector noises are uncorrelated. In the figure \ref{fig:outline}, we show a schematic outline of how the technique $\mu-$\texttt{GLANCE} is applied to the data.
We get the best-fit GW waveform given the unlensed hypothesis posterior distributions of all source intrinsic and extrinsic parameters of the specific GW event. We subtract this best-fit unlensed template from the data to obtain detector-specific residuals: $R_i(t) \equiv d_i(t) - h_{ BF,  i}(t)$, here $R_i(t)$ is the residual at the detector $i$, calculated by subtraction of the detector specific best-fit waveform $ h_{ BF, i}$ from the data $d_{i}(t)$. $R_{i}(t)$ may or may not contain microlensing characteristics. We expect that if a microlensing signature is present on top of the signal, it must be present in the residuals from all detectors. Thus to look for a common feature in the residual, we use cross-correlation on the time domain residuals from different detectors. The cross-correlation of the residuals between detectors $x$ and $x'$ is given by,
\begin{equation}\label{eq2}
    D_{x x'} (t) \equiv R_x \otimes R_{x'} = \frac{1}{\tau} \int_{t-\tau/2} ^{t+\tau/2} R_{x} (t') R_{x'} (t' + t_d) dt',   
\end{equation}
where, $R_i(t) = r_i(t) + n_i(t)$, here $r_i(t)$ denotes the residual GW, $n_i(t)$ denotes the noise of the detector and $t_d$ incorporates the time-delay between the arrival of the signal at different detectors due to the finite propagation speed of GWs. Here, we have chosen a cross-correlation timescale of $0.125s$. Note that, the choice of the cross-correlation timescale is made such that it is  sufficiently long to suppress the noise fluctuations, and also not too long to wash out the GW signal features. This timescale is typically chosen to be of the order or a little shorter that the signal duration in the observable frequency band.

\section{Application on up to O3 data to find lensed candidates}\label{sec3}

We have applied our technique on the LVK public data to look for potential microlensing candidates. Out of the total 90 confident GW events, we test the events that are observed with both L1 and H1 detectors \footnote{The Microlensing cross-correlation process requires the observability of an event in at least two detectors. Here, we do not consider any combination with the Virgo detector since its noise levels are higher than the two LIGO detectors.}. We also discarded \href{https://gwosc.org/eventapi/html/GWTC-1-confident/GW170817/v3/}{$\rm GW170817$} since the event luminosity distance was small enough (approximately $\approx 40$ Mpc) that there is hardly any required lensing optical depth along its path.

We select 72 GW events that are observed with both the LIGO observatories. For each selected event, the process to search for its microlensing signatures is shown by a schematic diagram in figure \ref{fig:outline}. In this work, we have chosen the \texttt{IMRPhenomXPHM} \citep{Pratten:2020ceb} waveform model for the analysis. Later on, a comparison between \texttt{IMRPhenomXPHM} and \texttt{SEOBNRv4PHM} \citep{Ossokine:2020kjp} waveforms has been performed to observe how much the results would have varied if we changed the waveform given the same set of source parameters \footnote{Putting the same values of parameters in different waveforms results in slightly different waveforms due to waveform modeling error \citep{Owen:2023mid}.}. We first use the unlensed posterior distribution of any specific event and take the maximum-likelihood values for all 15 GW parameters from the samples for the \texttt{IMRPhenomXPHM} waveform. By putting those parameters into the waveform models, we calculate the best-fit waveform model. To obtain the residuals, we subtract this best-fit waveform from the data. After whitening and frequency band-passing, we perform cross-correlation on the time-domain residuals from the detectors H1 and L1. We have shown it later that the waveform systematics is not a major source of uncertainty. However, to keep waveform error in check we use the piece of the waveform where the frequencies lies within $f_{\rm min}$ (Here we have chosen 20 Hz) and $f_{\rm ISCO}  = \frac{4.4 \rm kHz}{(m_1+m_2)}$ \footnote{$f_{\rm ISCO}$ is the frequency of the innermost stable circular orbit (ISCO) frequency for the compact binary.}. Since waveform systematic error increases with frequency, we select the part of the waveform up to $f_{\rm ISCO}$, where different waveforms tend to come in well agreement with one another. The lower end frequency cut-off $f_{\rm min}$ is chosen to be 20Hz to avoid contamination from waveform systematics due to effects such as eccentricity and low-frequency detector noise. Since the analysis is performed in the time-domain, we consider the portion of the residual cross-correlation over the time frame ($[t_{20}, t_{\rm ISCO}] $) here $t_{\rm min}$ corresponds to the GPS time when the dominant frequency of the signal is $f_{\rm min} =20$ Hz and similarly $t_{\rm ISCO}$ is the GPS time when the dominant frequency of the signal is $f_{\rm ISCO}$. Also, to compare the strength of the residual cross-correlation with the strength of the noise cross-correlation, we use a lensing signal-to-noise ratio to quantify the significance of any microlensing signature in the data with respect to the detector noise properties.

In order to check for possible contamination in the data, we make sure that the cross-correlation residual signal is present in the data for significant duration of time and not localized in a small fraction of the signal duration. Any microlensing/lensing feature, if present, is spread across the GW frequency range.  Therefore, any microlensing feature from the residual cross-correlation needs to be gradually building up with frequency, whereas any waveform systematics, detector glitch, non-lensing artifacts will appear like a sudden increase in the cross-correlation at the high frequencies. To understand whether the residual cross-correlation is non-localized and how the residual cross-correlation builds up, we take the addition of all the time-domain residual cross-correlation values over this $[t_{20}, t_{\rm ISCO}]$ period of time, given by (refer to \ref{eq2} for $D_{xx'}(t)$)
\begin{equation}\label{eq:sum_cc}
    S_{xx'}(t) =  \sum _ {t'=t_{20}} ^{t'=t\leq t_{\rm ISCO}} D_{xx'}(t')  \hspace{0.25cm},
\end{equation}
Here $S_{xx'}(t) $\footnote{We will use the notation $S_{xx'}$ to denote the cumulative signal in the paper to make the notation less cumbersome.}  takes the sum of all cross-correlation points between the detectors $x$ and $x'$ starting from $t_{20}$ up to the time $t \leq t_{\rm ISCO}$. We select the events which passes a significant amount of time through the $20Hz$ to $f_{\rm ISCO}$ frequencies because for short signals it is easy to get confused by waveform systematics and microlensing signatures just by looking at the residual cross-correlation. Therefore, if and only if the summation contains at least four bins i.e. the duration from $t_{20}$ to $t_{\rm ISCO}$ is more than or equal to $0.5s$ the residual features are considered as non-localized. The non-localized nature is very important to distinguish between microlensing modulation and waveform model based errors. More on this is discussed in selection criteria mentioned in the following subsection. The event's status for microlensing is checked only if it passes through the selection criteria. Note that, for high-mass binary black-hole (BBH) events the signals are shorter and if the duration $t_{\rm ISCO} - t_{20}$ is less than $0.5s$ one cannot conclude robustly whether any residual cross-correlation is because of systematics error or microlensing-like signatures in the data.

In order to quantify the statistical significance of the cross-correlation residual signal $S_{xx'}$, we compare it standard deviation of the time-summed noise cross-correlation $N_{xx'}$ defined as,
\begin{equation}
    N_{x x'} (t) \equiv n_x \otimes n_{x'} = \frac{1}{\tau} \int_{t-\tau/2} ^{t+\tau/2} n_{x} (t') n_{x'} (t') dt'.
\end{equation}
We used the standard H1 and L1 noise power spectral densities as a representative of noise characteristics for the specific observational run \footnote{H1- O1, O2, O3a, O3b chosen PSDs are taken from the following site \citep{H1_O1, H1_O2, H1_O3a, H1_O3b}. L1- O1, O2, O3a, O3b chosen PSDs are taken from the following sources \citep{L1_O1, L1_O2, L1_O3a, L1_O3b}.}. 
This PSD is used to simulate realistic noise and with similar post-processing on the noise as applied to the data, we calculate the noise cross-correlation. In the same duration $t_{\rm ISCO} - t_{20} = \Delta t$, we simulate time-stacked noise cross-correlation. Finally we estimate the standard deviation (denoted by $\sigma_{xx'}$) of the these summed noise cross-correlation over many different noise realizations using the relation
\begin{equation}
    \sigma_{xx'} = \sqrt{ \textbf{Var} \left(\sum_{t=0} ^{t=\Delta t} N_{xx'} (t)\right) } \hspace{0.25cm},
\end{equation}
here we denote the variance of the noise cross-correlation residual by \textbf{Var}.
Using the above defined signal cross-correlation and noise cross-correlation, we define the residual cross-correlation signal-to-noise ratio (SNR) as
\begin{equation}\label{eq:snr}
    \rho_{\rm residual} \equiv \frac{S_{xx'}}{\sigma_{xx'}} = \frac{  \sum _ {t_{20}} ^{ \leq t_{\rm ISCO}} D_{xx'}}{\sqrt{ \textbf{Var} \left(\sum_{0} ^{\Delta t} N_{xx'} \right) }} \hspace{0.25cm}.
\end{equation}

The quantity $\rho_{\rm residual}$ captures the deviation of the signal strength as compared to the typical strength of the noise cross-correlation fluctuations. It is important to mention that for data and noise the cross-correlation timescale is the same, here we have chosen it to be $0.125s$. The SNR can also be mapped to false alarm rate (FAR) in a straightforward way. However, we prefer to quote the results in terms of SNR for quantifying the significance and use an SNR threshold of three as a selection criterion, which is commonly used in astrophysics. For an application of the residual cross-correlation technique with FAR in consideration, you may look at \citep{Chakraborty:2024mbr}.

\subsection{Microlensing detection and Classification}\label{subsec:detect}
\begin{figure}
    \centering    \includegraphics[width=\linewidth]{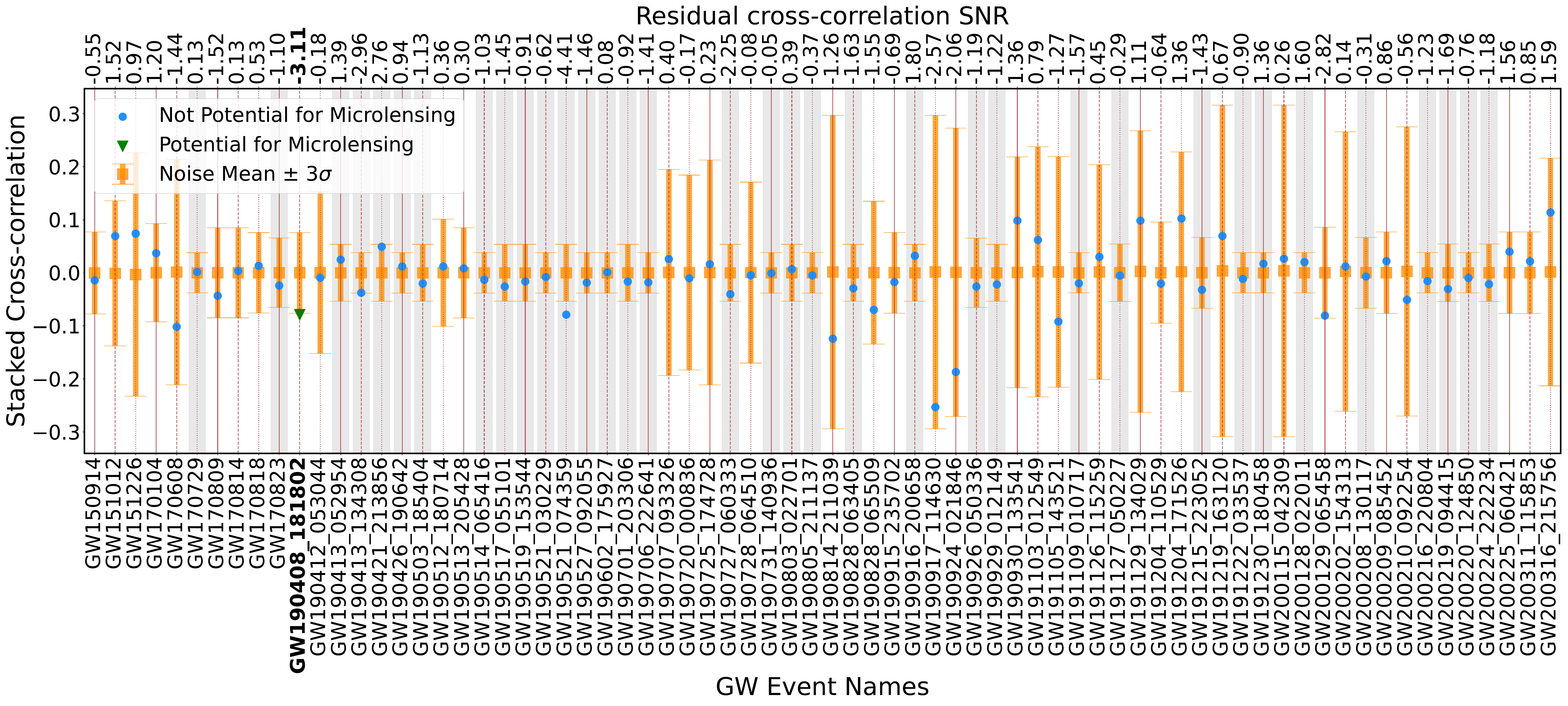}
    \caption{The figure shows the residual cross-correlation SNRs for each selected event. The orange error bars represent the noise characteristics $\sigma_{xx'}$, the blue dots represents the signal strength $S_{xx'}$ from the same equation and the gray band represents the rejection band, which discards events for having $< 0.5s$ inspiral duration. The green triangle shows the selected event, which is considered as a microlensed GW candidate.} 
    \label{fig:all_events}
\end{figure}

By using the technique discussed in the previous section, we apply the lensing search technique $\mu-$\texttt{GLANCE} on the GW events up to GWTC-3 which are detected with a network matched filtering SNR $8$ and above. 
In the figure \ref{fig:all_events}, we summarize the results of all the events with an matched filtering SNR at least 8. The events are labeled along the x-axis. The Signal from the cross-correlation $S_{xx'}$ is plotted with blue dots in the y-axis along with error bars coming from $\sigma_{xx'}$. We apply vertical gray bands to those events which has $t_{\rm ISCO} - t_{20} < 0.5s$. The event which is interesting for microlensing is shown with green triangle. The residual cross-correlation SNR $\rho_{\rm residual}$ is mentioned on top for each event.

A detailed version results of the search are shown in the figures \ref{fig:f1} to \ref{fig:f4} of appendix \ref{app:a}, in a chronological sequence. For a certain row in these four figures, the first panel (from left to right) shows the post-processed residual at the H1 detector, the second panel shows the residual at the L1 detector. The third panel shows the cross-correlation between the two detector residuals. The fourth panel depicts the gradual growth of the cross-correlation as the time progresses. To minimize effects due to the waveform systematics, we only focus the time interval between which the signal lies between, $20$Hz and $f_{\rm ISCO}$. Here in each panel, the navy blue dotted line represents the GPS time when the signal has the $20$Hz as its most-dominant component, and the dark-red dotted line which shows the GPS time when $f_{\rm ISCO}$ is the most dominant frequency in the GW signal. The black dotted line further shows the GPS time of the coalescence. The residual cross-correlation SNR, defined as in equation \ref{eq:snr} is mentioned on the top of the fourth panel for every row. 

We then apply the selection criteria to mitigate the contamination from noise and systematics uncertainty on the microlensing searches. We list down below the criteria used in the analysis :

\begin{enumerate}
    \item \textbf{Check for the non-localized nature of the residual cross-correlation:}  To detect microlensing features from the GW signal, we choose to the restrict the analysis up to the inspiral phase since the merger-ringdown phases are more likely to be contaminated by waveform systematic. Further, we understand that a truly microlensing feature, if present, must be spread across all the frequencies of the GW. Therefore, a gradual growth in the cumulative residual cross-correlation confirms the presence of an unmodeled signal beyond the modeled GW signal. The cumulative residual cross-correlation demonstrates whether the residual cross-correlation is localized or spread throughout the signal. Noise in the detector can produce jitters in the cross-correlation, but these fluctuation are not persistent and therefore die down over time. The persistent non-localized nature of the residual cross-correlation helps us quantify the degree of faith for a detection of a potentially microlensed GW candidate. Thus to deal with the waveform systematics and jitters because of noise and to observe the growth of the residual cross-correlation over the inspiral phase, we select those events if their inspiral duration is at least $0.5s$. The events which do not pass this criteria are shown with vertical gray band in the figure \ref{fig:all_events}.
    
    \item \textbf{Negative cross-correlation in the H1 and L1 detectors:} Due to the specific orientation of the LIGO Hanford Observatory arms with respect to the LIGO Livingston Observatory arms \footnote{Please refer to the article \citep{lhantenna, det_antenna} for more details on the detector antenna patterns.}, the residual cross-correlations for any commonalities present in the residuals from these two detectors shows a negative value. Thus we select the events which shows a residual cross-correlation SNR of negative parity. If and only if the cross-correlation keeps on building in the negative direction, we consider that event for microlensing candidature. The events which pass this criteria, has a negative residual cross-correlation SNR as mentioned at the top x-axis of the figure \ref{fig:all_events}.
    
    \item \textbf{Strength of the cross-correlation signal:} To estimate the strength of the residual cross-correlation, we compare its strength with respect to the noise cross-correlation in term of the residual cross-correlation SNR as given by equation \ref{eq:snr}. Here we choose the residual cross-correlation SNR threshold to be at $|\rho_{\rm residual} ^{\rm th}|= $3. The 3-sigma condition sets the strength of the residual cross-correlation more than 99\% away from underlying noise-statistics, which helps to filter out only the interesting events with a common residual to scrutinize them further for microlensing. Imposing a more stringent cut-off condition, defies the purpose of detecting a microlensing signature from the data, where residual can be an order of magnitude smaller than the strain signal.\footnote{In future for high SNR GW events, use of higher threshold such as $5\sigma$ will be appropriate to claim a discovery of lensed GW signal.} The events which qualify this criterion along with the previous two is highlighted in bold fonts and a green triangle marker is used for them. 
\end{enumerate}


With these selection criteria, we observe that only one event \href{https://gwosc.org/eventapi/html/GWTC-2/GW190408_181802/v1/}{{GW190408\_181802}} is able to pass as a candidate for microlensing. The table \ref{tab:snr_table}, summarizes the details of the events with the residual cross-correlation SNR significance in descending order.

\begin{figure}
    \centering
    \includegraphics[width=\linewidth]{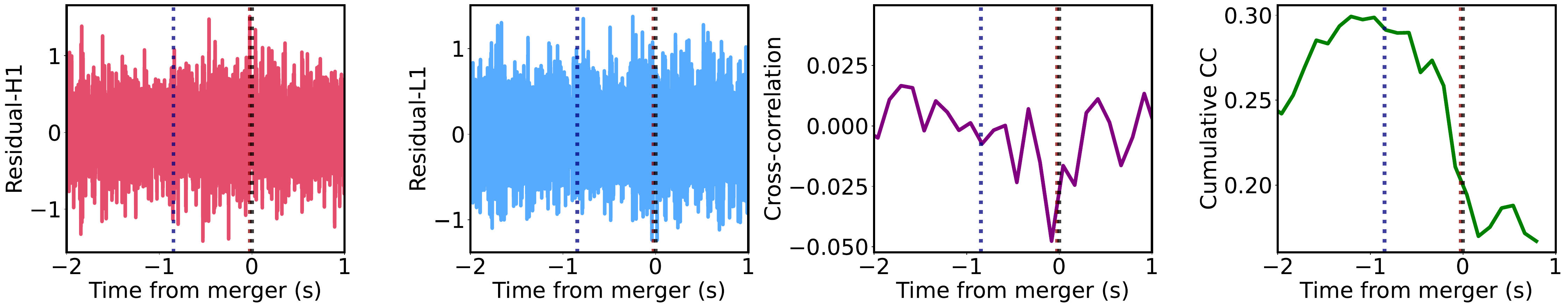}
    \caption{The figure shows the details of the event $\rm GW190408\_181802$, the only event to satisfy all the mentioned conditions to pass the microlensing search test through cross-correlation. The first panel (from left to right) shows the residual at the Hanford detector, the second panel shows the residual at the Livingston detector. The third panel depicts the residual cross-correlation between the two detector residuals with a cross-correlation timescale of $0.125s$. The horizontal band shows the standard-deviation of the noise cross-correlations. The fourth panel shows the build-up of the cross-correlation over time. We compare the strength of the build-up (over the aforementioned frequency range) with noise cross-correlation build-up, which provides a signal-to-noise ratio of 3.11}
    \label{fig:GW190408}
\end{figure}

\begin{table}[h!]
\centering
\begin{tabular}{lrrrrrc}
\toprule
\textbf{Event Name} & \boldmath{$m_1 (M_{\odot})$} & \boldmath{$m_2 (M_{\odot})$} & \textbf{Distance (Mpc)} & \textbf{$\rho_{\rm residual} \geq \rho_{\rm th} = 3$} & \textbf{Localized} & \textbf{Selection Criteria Matched} \\
\midrule
\rowcolor{gray!20} 
\href{https://gwosc.org/eventapi/html/GWTC-2/GW190408_181802/v1/}{\textbf{GW190408\_181802}}  & \textbf{34.551} & \textbf{22.721} & \textbf{1392.358} & \textbf{(-)3.111, Yes} & \textbf{No} & \textbf{Yes} \\  
\href{https://gwosc.org/eventapi/html/GWTC-3-confident/GW200129_065458/v1/}{GW200129\_065458} & 51.639 & 20.746 & 791.314 & (-)2.822, No & No & No\\  
\href{https://gwosc.org/eventapi/html/GWTC-2.1-confident/GW190917_114630/v1/}{GW190917\_114630}  & 11.535 & 2.276 & 484.152 & (-)2.568, No & No & No\\  
\bottomrule
\end{tabular}
\caption{Residual cross-correlation SNR values and classification of GW events into events interesting for microlensing. We list the maximum-likelihood values of the detector-frame masses, distances, and cross-correlation based residual cross-correlation SNR values for each event. We then mention whether the signal is very localized to be tested. Finally if all three aforesaid conditions are met, the event is classified as an interesting event for microlensing.}
\label{tab:snr_table}
\end{table}

We observe that the event $\rm GW190408\_181802$ has a residual cross-correlation SNR marginally above (-)3 and the signal is non-localized. The residual, cross-correlation and its non-localized build-up in the $[t_{\rm 20}, t_{\rm ISCO}]$ timescale is shown in the figure \ref{fig:GW190408} which is greater than $0.5s$. The choice of \texttt{SEOBNRv4PHM} waveform model and its effect on the residual cross-correlation for the event is shown in the next subsection.
Thus it shows that $\mu$-\texttt{GLANCE} can perform a unmodelled cross-correlation search through the events to look for the presence of any microlensing features. It does not rely on a spherical mass distribution of the lens system. The effects of waveform based error and noise based error and their relative strengths are compared in the next subsection \ref{subsec:systematics}. Although individual detector noise gets suppressed in cross-correlation, waveform systematics error remains. However, in the next subsection, we discuss that the waveform systematics error is around 5-9 times smaller than the noise-based error. However, cross-correlation noise error cannot affect the residual cross-correlation since the detector noises are uncorrelated. This implies that the waveform error affects the cross-correlation based results much more than what noise error can because the waveform based errors are common to different detectors.

Through these tests we confirm only one potential interesting candidate for microlensing, which can be further scrutinized according to the flowchart shown in figure \ref{fig:outline}. The event $\rm GW190408\_181802$ is the only event which passes through the aforesaid selection criteria, signifying the robust detection capability of $\mu$-\texttt{GLANCE} since the method did not pick up a large number of events potential for microlensing checks. The event was not flagged as important by any of the previous work where joint parameter space exploration of the source and the lensing parameters was mostly performed assuming spherical microlensing models, a few of such lenses and their effects can be found here \citep{Takahashi:2003ix}. We also emphasize that if the event is microlensed its microlensing features should be at the level of characterization along with a significant residual cross-correlation. In the subsection \ref{subsec:PE}, we have performed a Bayesian analysis to estimate the evidence of any microlensing signature from the event.

\subsection{Systematics on waveform}\label{subsec:systematics}
\begin{figure}
    \centering
    \includegraphics[width=\linewidth]{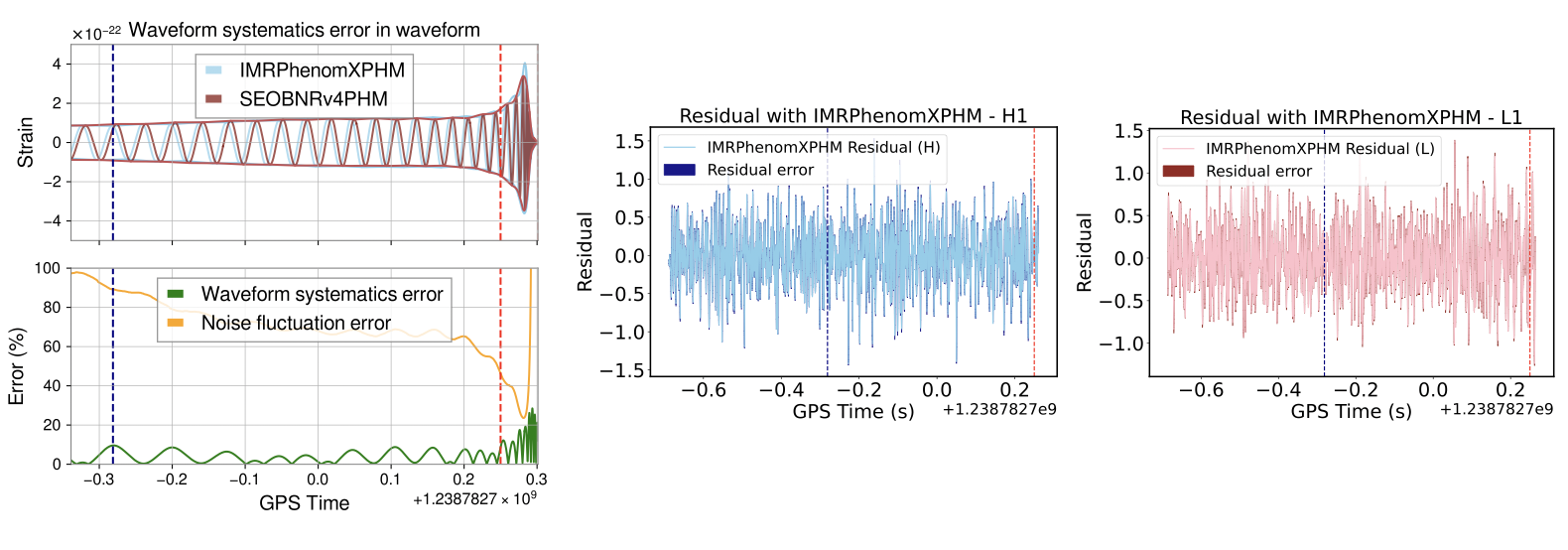}
    \caption{The leftmost top panel of the figure shows the comparison of the time domain waveform for \texttt{IMRPhenomXPHM} and \texttt{SEOBNRv4PHM}. Bottom of that, we show the comparison between the waveform error (in green) and noise error (in orange). We observe that within the frequency range, the waveforms are agreeing well to each other. The mismatch at any time is at max 10\%, therefore the residuals are affected at max by this amount. We also note that, after $t_{\rm ISCO}$ the error in the waveform grows rapidly, it becomes 30\% close to the merger. This proves the benefits of the requirement of the waveform up to $f_{\rm ISCO}$ to mitigate any waveform model based error. The next two panels on the right show the residual for the GW event $\rm GW 190408\_181802$ with \texttt{IMRPhenomXPHM} waveforms along with their systematic error for detectors H1 and L1. For all these figures we consider the region between the navy blue and the red line denoting the frequencies 20Hz and $f_{ISCO}$ respectively. The waveform based error propagates to the residuals which in turn affects the residual cross-correlation. However, the residual error from systematics being less than or equal to 10\%, it will not significantly impact the results obtained through cross-correlation.}
    \label{fig:systematics}
\end{figure}

The residual cross-correlation depends on the calculation of the residuals, which for different waveform models vary slightly even when the source parameters are kept the same. Thus a residual cross-correlation with a particular waveform model, may be stronger/weaker than with the choice of a different waveform model. Thus to minimize such waveform systematics effect \citep{Read:2023hkv} on the residual cross-correlation, we have only chosen the the inspiral phase of the GW for the residual cross-correlations, up to the frequency of $f_{\rm ISCO}$, which contains maximum information about the nature of the GW (since this phase contains the maximum amount of the matched-filtering SNR of an event). Up to this frequency, all waveform models tend to agree well with each other. However, this is not the case for the merger and the ringdown phase, where different waveforms start to disagree with one another. Thus for the sake of unanimous acceptance of the results, we omit any contribution from the merger and ringdown phases in our work. The waveform systematics can affect the detection not only for a microlensing detection but also hampers the detection of strongly lensed GW signals \citep{Garron:2023gvd, Keitel:2024brp}. In the top plot of the leftmost panel of the figure \ref{fig:systematics}, we present the comparison between the \texttt{IMRPhenomXPHM} signal and the  \texttt{SEOBNRv4PHM} signal given the \texttt{IMRPhenomXPHM} set of best-fit parameters from the parameter estimation for the event $\rm GW190408\_181802$. The errors due to the mismatch of the two waveforms are shown in the panel below that. The green curve shows that the systematics errors are of the order of $1- 10\%$ within the time interval $[t_{20}, t_{\rm ISCO}]$, implies a mild shift of the residual cross-correlation SNR, when we change waveforms. In comparison the orange curve shows that the errors due to noise is larger by almost a factor of 9 at $t_{20}$ and a factor of 5 at $t_{\rm ISCO}$ than the systematics error. We also note that, the systematics error grows rapidly after the $f_{\rm ISCO}$, from 10\% at $t_{\rm ISCO}$ to 30\% at $t_{\rm merger}$, where $t_{\rm merger}$ is the GPS time of merger and corresponds to the peak of the strain. This highlights that the waveform error can be minimized only by rejecting the contribution from higher frequencies of the GW. Next in the two panels on the right side, we show the whitened residuals with the H1 and L1 detectors using the \texttt{IMRPhenomXPHM} waveform. In both of those panels, we have also shown the the error in the residual evaluation due to the waveform systematics errors. We previously observed that the error in the waveform due to waveform model systematics is around 10\%. This error propagates to the residual calculation and thus its estimation is subjected to an error of around 10\%. Although the strength of the noise error is larger than the systematics based error, we would like to emphasize that, the error due to noise is not correlated across detectors and thus it does not propagates to the residual cross-correlation. In contrary the waveform based error being common across detectors propagates to residual cross-correlation. Thus microlensing searches based on residual cross-correlation can be affected by the waveform systematics. This effect can be minimized by considering the performance of the technique within the [$t_{20}, t_{\rm ISCO}$] time interval. Thus, all events are assessed for microlensing based on the time interval spent by the signal within the frequencies $20$ Hz-$f_{\rm ISCO}$. 

\subsection{Characterization of the microlensing parameters}\label{subsec:PE}

Microlensing causes a frequency-dependent amplitude and phase modulation on the GW waveform. To understand whether the microlensing features are present on the interesting event we have, we assess the strength of the lensing amplification imprinted on the GW. We have already performed an unmodelled search through cross-correlation so far and found one interesting event, but if microlensing features are present in the GW, it needs to detected and characterized. If the support towards the microlensing parameters is low, we cannot call the event as a potential candidate for microlensing. Therefore to characterize the effects due to microlensing, we perform a Bayesian analysis on the lensing parameters.  We note that, microlensing effects have no strong impact on the chirping behavior ($\dot{f}/f$) of the GW and so does not impact the chirp mass inference significantly. As shown in our previous work $\mu$-\texttt{GLANCE} \citep{Chakraborty:2024mbr} from mock data, the microlensing modulations are nearly non-degenerate with the masses and also the aligned spins. Even if the correlation between source and lensing parameters would have existed, the marginal significance of $\rm GW190408\_181802$ as the candidate passing through the selection criteria, implies that if the lensing parameters are present their presence is very weak and the measurement will not impact the inference of the source properties. Thus we perform an estimate only on the lensing parameters keeping the source parameters fixed at their best-fit values to reduce the computational cost. 

We perform the Bayesian analysis on the microlensing regime using equation \eqref{eq:lens}. The parameter $k$ which allows the transition from the wave optics to the ray optics is vanishingly small for the short length of this signal and therefore may be ignored for this analysis. We note that any effects of strong lensing goes into the magnification term $a$, which in turn is absorbed in the luminosity distance estimation \footnote{The strong lensing phase shift effect is analogous to effect of a different coalescence phase ($\phi$)}. Thus we have a set of three effective parameters $b$, $f_0$ and $\phi_0$ to capture the microlensing signatures. The priors on these parameters are chosen to be,
\begin{table}[h!]
\centering
\begin{tabular}{cccc}
\hline
\textbf{Parameter} & \textbf{Distribution Type} & \textbf{Minimum} & \textbf{Maximum} \\ \hline
$b$ & Uniform & 0 & 1 \\ \hline
$f_0$ & Uniform & 5 Hz & $f_{\rm ISCO}= 76.8$ Hz \\ \hline
$\phi_0$ & Uniform & 0 & $2 \pi$ \\ \hline
\end{tabular}
\caption{Prior distributions for the microlensing parameters} 
\label{table:priors}
\end{table}

\begin{figure}
    \centering
    \includegraphics[width=0.95\linewidth]{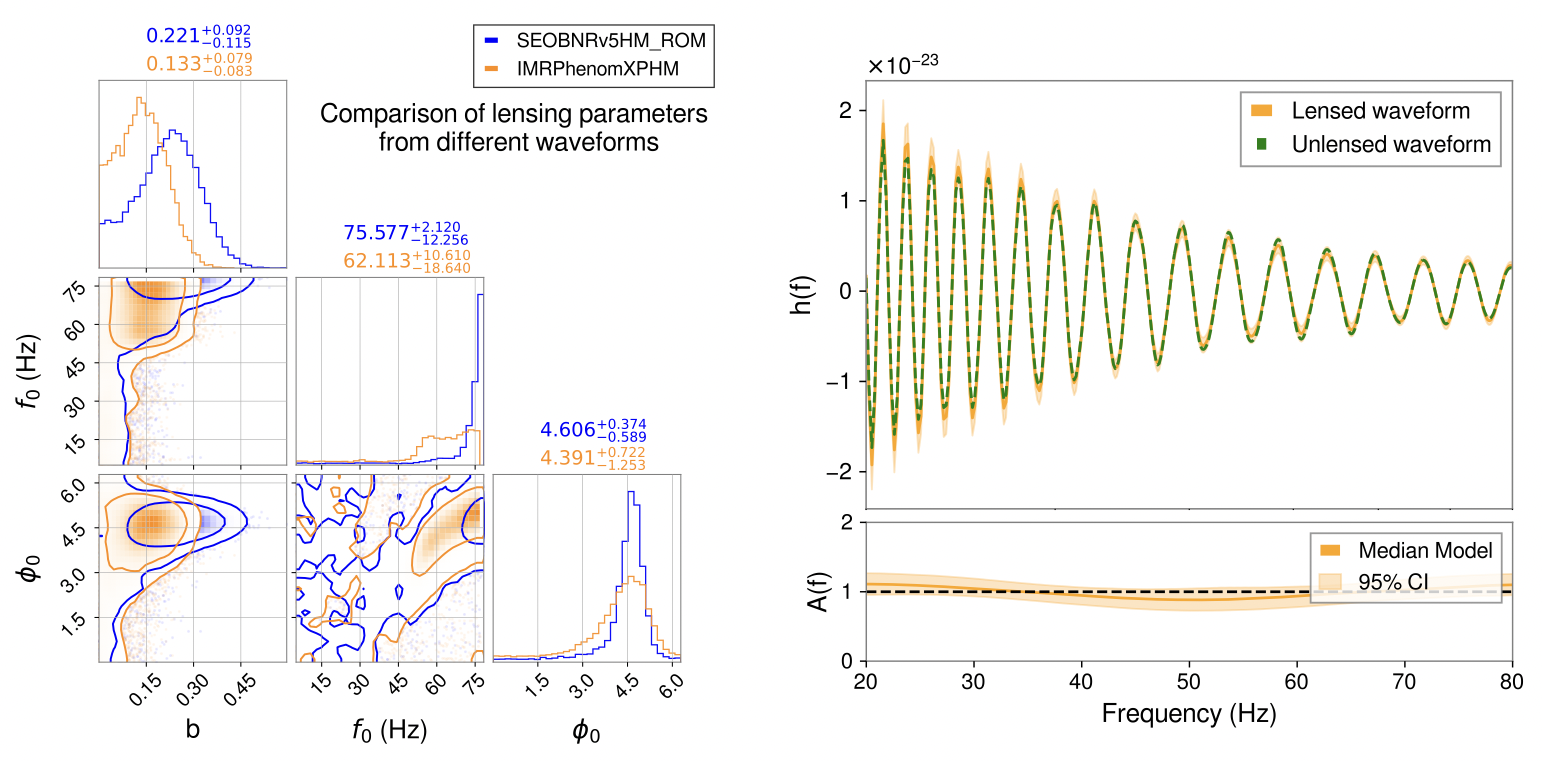}
    \caption{The figure show the parameter space exploration for the lensing parameters. The left panel shows the estimation of the lensing parameters in a completely microlensing regime ($k=0$). The contours show 1-$\sigma$ and 2-$\sigma$ contours around the medians. On top of the posteriors, the median value and the uncertainty (1-$\sigma$) in its measurement is shown. We compare the posteriors of the lensing parameters for \texttt{IMRPhenomXPHM} and \texttt{SEOBNRv5HM\_ROM} waveforms. It shows the posteriors are in very good agreement with the the fact that the presence of microlensing features on the signal is weak. The waveform from the lensed waveform is compared with the unlensed waveform with best-fit lensing and source parameters is the right top panel. We see small wavy nature of the lensed waveform around the unlensed. However, the strength of the microlensing is not significant and the railing of the $f_0$ posterior suggests a modulation frequencies are larger, pointing towards a static nature of the amplification. Below the panel, we have shown the frequency dependent amplification factor estimated from the event data and its variation with frequency.}
    \label{fig:PE_190408}
\end{figure}

We choose a Gaussian likelihood of the form 
\begin{equation}
\log(L) = -\sum_{x}\sum_{f=f_{\rm low}} ^{f=f_{\rm high}} \left[ \frac{|d_{x}(f) - h_{x}^{lensed} (f; \{ \theta \})|^2}{2 C_{x}(f) } + \frac{1}{2}\log(2 \pi C_{x}(f)) \right] ,
\end{equation}
where $d_x(f)$ is the data and $C_{x}(f)$ is the noise covariance of a detector (denoted by $x$). The lensed waveform model is: $h_{x}^{\rm lensed} (f; \{ \theta \}) = h^{\rm unlensed}_x(f) \times A(f; \{\theta\})$. We have used Metropolis-Hastings Markov Chain Monte-Carlo (MCMC) sampler using the tool \texttt{emcee: The MCMC Hammer} \citep{Foreman_Mackey_2013} and plotted the joint distributions using \texttt{CORNER} \citep{corner} in the figure \ref{fig:PE_190408}. The left panel of the figure shows the exploration of the lensing parameter space. The results obtained are for both choices of waveform models: \texttt{IMRPhenomXPHM} and \texttt{SEOBNRv5HM\_ROM} are well in agreement.  The MCMC result shows that the value of the parameter which regulates the  strength of the microlensing signature is $b = 0.133 ^{+0.079} _{-0.083}$ for the \texttt{IMRPhenomXPHM} waveform model and $b = 0.221 ^{+0.092} _{-0.115}$ for the \texttt{SEOBNRv5HM\_ROM} waveform model. A small $b$ value suggests no strong evidence of microlensing like features in the waveform. Also, the characteristic oscillation frequency of microlensing denoted by $f_0$ exhibit a large value for both the waveform models, indicating only that the residual signal can only support weak modulation. The initial phases do agree as well for both waveform. Overall, in brief, the above microlensing estimations for different waveforms are in accordance and their values suggest no strong presence of a microlensing feature in the data. 

To validate our conclusion even further, we plot in figure \ref{fig:PE_190408} right-top, panel the \texttt{IMRPhenomXPHM} waveforms for the no-lensing hypothesis and with the lensing hypothesis with the medians of the posteriors of the lensing parameters (b, $f_0$, and $\phi_0$). The amplitude modulation due to microlensing are small enough to make the lensed and the unlensed waveform to overlap up to a very high degree. The bottom right panel of the same figure shows the amplitude $A(f)$ modulation with the variation in frequency, with the set of median values of the lensing parameters with their associated $95\%$ credible interval uncertainties. This amplification is obtained from the data and is the same as the one shown on the top panel lensed waveform. It shows the tiny deviations from line $A(f) = 1$ caused by the estimated lensing amplification curve from that event, indicating no statistically significant feature of non-zero microlensing signal. So, based on this result we conclude that there is no strong evidence of microlensed event up to GWTC-3.   

\section{Comparison with the previous microlensing searches and the necessity of $\mu$-\texttt{GLANCE} as a search technique}\label{sec1}
From the data up to the third observational run, the LVK collaboration searched for potential microlensing signatures on the GW events for simple lens mass model and without including both microlensing and strong lensing scenario \citep{LIGOScientific:2021izm, LIGOScientific:2023bwz}. This implements a similar approach to the work \citep{Wright:2021cbn} where microlensing parameters (mass of the lens, impact parameter) are estimated alongside the source parameters (chirp mass, mass-ratio). The joint estimation helps to constrain any  wave-optics lensing effects imprinted on the GW by a spherically symmetric lens model. By calculating the Bayes factor supporting the lensing hypothesis as shown in the work \citep{LIGOScientific:2023bwz}, the work finds no evidence of a microlensing candidate, with a maximum Bayes factor being $\rm log_{10} B^{microlensed}_{unlensed} = 0.799$ for the event \href{https://gwosc.org/eventapi/html/GWTC-3-confident/GW200208_130117/v1/}{$\rm GW200208\_130117$}. Machine-learning-based techniques \citep{Kim:2022lex} have also been applied to the O1-O2 data to search for microlensing effects present in the data. Through the spectrogram (q-transformation of the time-domain data, which maps the evolution of the GW event through the time-frequency bins) analysis, this method looks for subtle microlensing beating patterns present in the GW signal. It finds an interesting event \href{https://gwosc.org/eventapi/html/GWTC-2/GW190707_093326/v1/}{$\rm GW190707\_093326$}; however, with the p-value of the event being $p \geq 0.05$, the event is not very likely to be a lensed event. 

A few limitations of the previous works are that those works search for the point mass and other spherically symmetric lens effects on the GW signal, which is not the most generic type of lens. The deviation from the spherical geometry of a lens is the most likely physical situation for an astrophysical object and therefore the assumption of a lensing model may not be reasonable enough for a lensing detection technique. Also, the analysis techniques are susceptible to fluctuation in the strain data due to noise artifacts. Therefore, a noise artifact present together with the signal may be attributed falsely as microlensed due to its similarity with microlensed-like features. Furthermore, the effects of waveform systematics-based errors are not mitigated in the aforementioned works. The high-frequency regime ($f\geq f_{\rm ISCO}$), where the agreement between the different waveform models starts to fall apart, is a must that needs to be accounted for a robust detection algorithm.

The analysis presented in this work using $\mu$-GLANCE addresses these limitations by searching for microlensing signatures in a model-independent way using a cross-correlation technique which mitigates the effects from uncorrelated detector noise and uncommon systematics and takes into account the uncertainty due to waveform systematics. This has helped us in a robust analysis of microlensing from the GW data and rule out the evidence of any statistically significant microlensing signal. For the events $\rm GW190707\_093326$ and $\rm GW200208\_130117$, as we can mitigate the noise contamination based on the residual cross-correlation, and find no evidence of lensing (see figure \ref{fig:all_events}). So, the low-significant microlensing candidate events picked up by the previous works have naturally been ruled out here implying the robustness and efficiency of this work reflected through the high rejection standards. We found, in fact, one interesting candidate event in the residual test $\rm GW190408\_1818102$, which after the lensing parameter estimation shows no evidence of a microlensing feature. Hence, this analysis confidently rules out the presence of any statistically significant lensing signal from the GW catalog for the 72 events that are detected in more than one GW detector.

\section{Conclusions and Future Prospects}\label{sec4}

In this work, we conduct the first model-independent approach of deploying a residual cross-correlation-based microlensing search technique $\mu$-\texttt{GLANCE} on the public LVK data. The relative strength of the residual cross-correlation and the noise cross-correlation help us qualify an event as an interesting event for microlensing only if the residual cross-correlation is non-localized. We minimize waveform systematics-based errors in the calculation of the residual by trusting the results within the frequency range [20Hz, $f_{\rm ISCO}$]. Based on the residual cross-correlation strength compared to noise, we find an interesting event $\rm GW190408\_181802$.  However, no strong support for lensing parameters was found through lensing parameter space exploration in a Bayesian framework, which is consistent with the event's marginal significance. Therefore based on the two-step microlensing authentication, we reject the microlensing hypothesis $\rm GW190408\_181802$ of the event. We point out that in the previous works, the events ($\rm GW190707\_093326$ and $\rm GW200208\_130117$) which were found to be interesting microlensing candidates are not found to be passing through the cross-correlation-based selection because of robust detection technique and capability in mitigating noise. The only event $\rm GW190408\_181802$ which was selected as an interesting event by the cross-correlation search, is rejected being microlensed, based on non-evidence of strong support for the lensing parameters in the Bayesian analysis. This shows the robustness and efficiency of the method $\mu$-\texttt{GLANCE} through its high rejection ratio which allows for the first time to rule out the evidence of any statistically significant microlensed event in a model-independent way. 

The use of the $f_{\rm ISCO}$ as an upper limit in our analysis is intended to reduce uncertainties arising from waveform systematics when inferring microlensing signatures. While this condition is not inherently required for the cross-correlation technique to identify lensing, we observe that discrepancies between different waveform models begin to exceed 10\% beyond this frequency threshold (see figure \ref{fig:systematics} for the event GW190408\_181802). As a result we mitigate this enhanced waveform uncertainty to impact the lensing inference. This choice reduces the capability of the detection of lensing identification for sources whose signal is dominated by the merger-ringdown phases—particularly those with high redshifted chirp mass. In future, accurate modeling of waveform will be crucial to accurately capture microlensing signatures from a lensed GW signal.

In the future, with O5 detector sensitivities of the LVK detectors, and with the operation of LIGO-Aundha \citep{LIGO_India} in the upcoming years, we can perform a total of $^5C_2 =10 $ cross-correlations between the LIGO-Virgo-KAGRA detector pairs. This would allow robust detection and classification of the lensing signal by improving the residual calculation in $\mu-$\texttt{GLANCE}.  This will lead to a robust search for microlensing signatures from the data and open the road for the discovery of the first microlensed GW signal. The discovery space will grow significantly with the operation of the next generation GW detectors such as Cosmic Explorer \citep{Evans:2021gyd} and Einstein Telescope \citep{Punturo:2010zz, Maggiore:2019uih}. 


\section{Acknowledgments}
The authors are thankful to Prasia Pankunni for reviewing the manuscript
during the LSC Publications and Presentations procedure and providing useful comments. The authors express their gratitude to the \texttt{⟨data|theory⟩ Universe-Lab} group members for useful suggestions. This work is part of the \texttt{⟨data|theory⟩ Universe-Lab}, supported by TIFR and the Department of Atomic Energy, Government of India. The authors express gratitude to the computer cluster of \texttt{⟨data|theory⟩ Universe-Lab} for computing resources used in this analysis. We thank the LIGO-Virgo-KAGRA Scientific Collaboration for providing noise curves. LIGO, funded by the U.S. National Science Foundation (NSF), and Virgo, supported by the French CNRS, Italian INFN, and Dutch Nikhef, along with contributions from Polish and Hungarian institutes. The research leverages data and software from the Gravitational Wave Open Science Center, a service provided by LIGO Laboratory, the LIGO Scientific Collaboration, Virgo Collaboration, and KAGRA. Advanced LIGO's construction and operation receive support from STFC of the UK, Max-Planck Society (MPS), and the State of Niedersachsen/Germany, with additional backing from the Australian Research Council. Virgo, affiliated with the European Gravitational Observatory (EGO), secures funding through contributions from various European institutions. Meanwhile, KAGRA's construction and operation are funded by MEXT, JSPS, NRF, MSIT, AS, and MoST. This material is based upon work supported by NSF’s LIGO Laboratory which is a major facility fully funded by the National Science Foundation. We acknowledge the use of the following python packages in this work: NUMPY \citep{harris2020array}, SCIPY \citep{2020SciPy-NMeth}, MATPLOTLIB \citep{Hunter:2007}, PYCBC \citep{alex_nitz_2024_10473621}, GWPY \citep{gwpy}, LALSUITE \citep{lalsuite}, emcee: The MCMC Hammer \citep{Foreman_Mackey_2013} and CORNER \citep{corner}.

\bibliography{bibliography}{}
\bibliographystyle{aasjournal}

\appendix

\section{Appendix A: Full analysis results on the selected GW events}\label{app:a}

\begin{figure}
    \centering
    \includegraphics[width=0.9\linewidth]{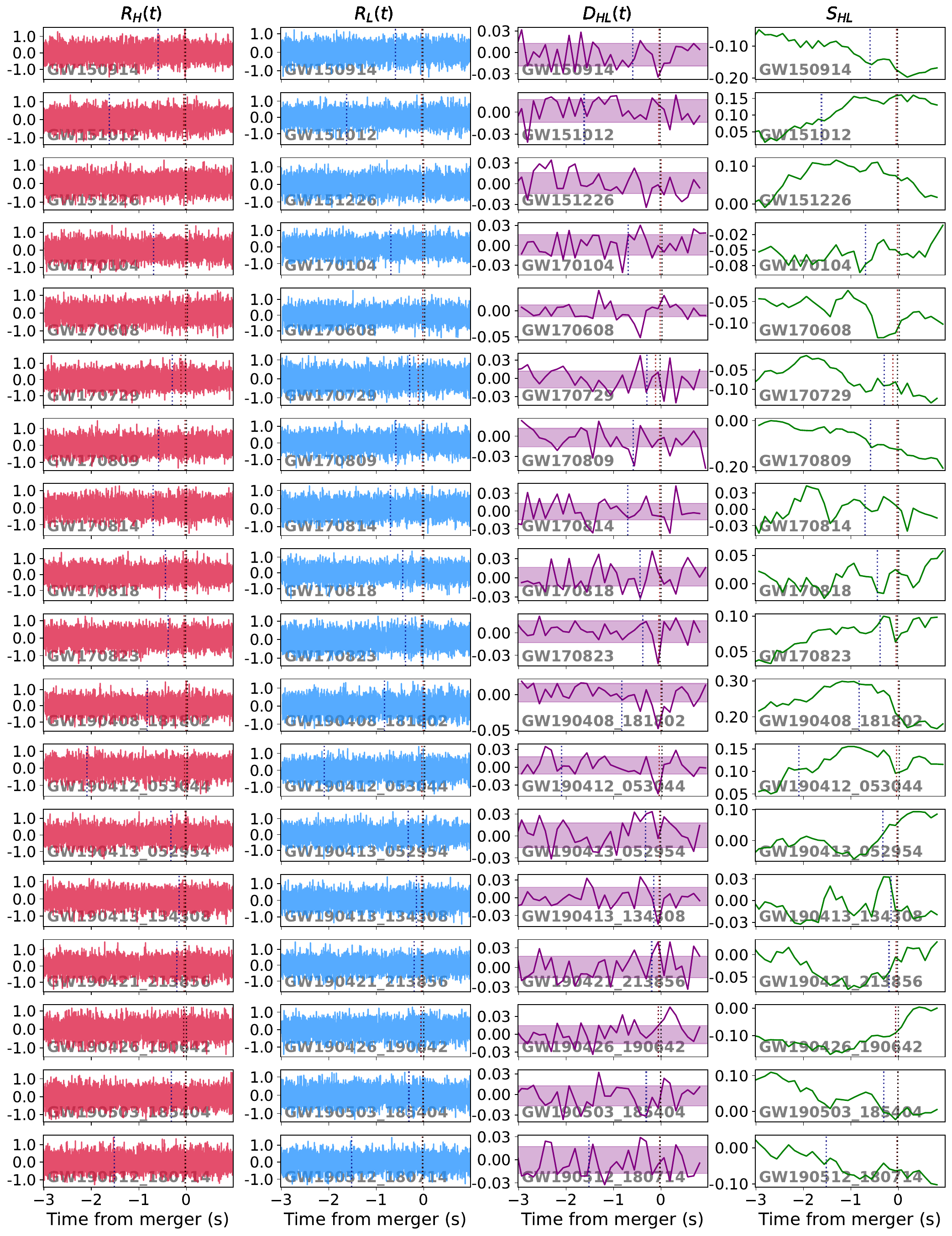}
    \caption{Performing $\mu$-\texttt{GLANCE} on the GW events dated between September 14th, 2015 to May 12th, 2019. For each row, the panels represent the residual from H1, the residual from L1, the cross-correlation of these two residuals and the cumulative changes of the cross-correlation.}
    \label{fig:f1}
\end{figure}

\begin{figure}
    \centering
    \includegraphics[width=0.9\linewidth]{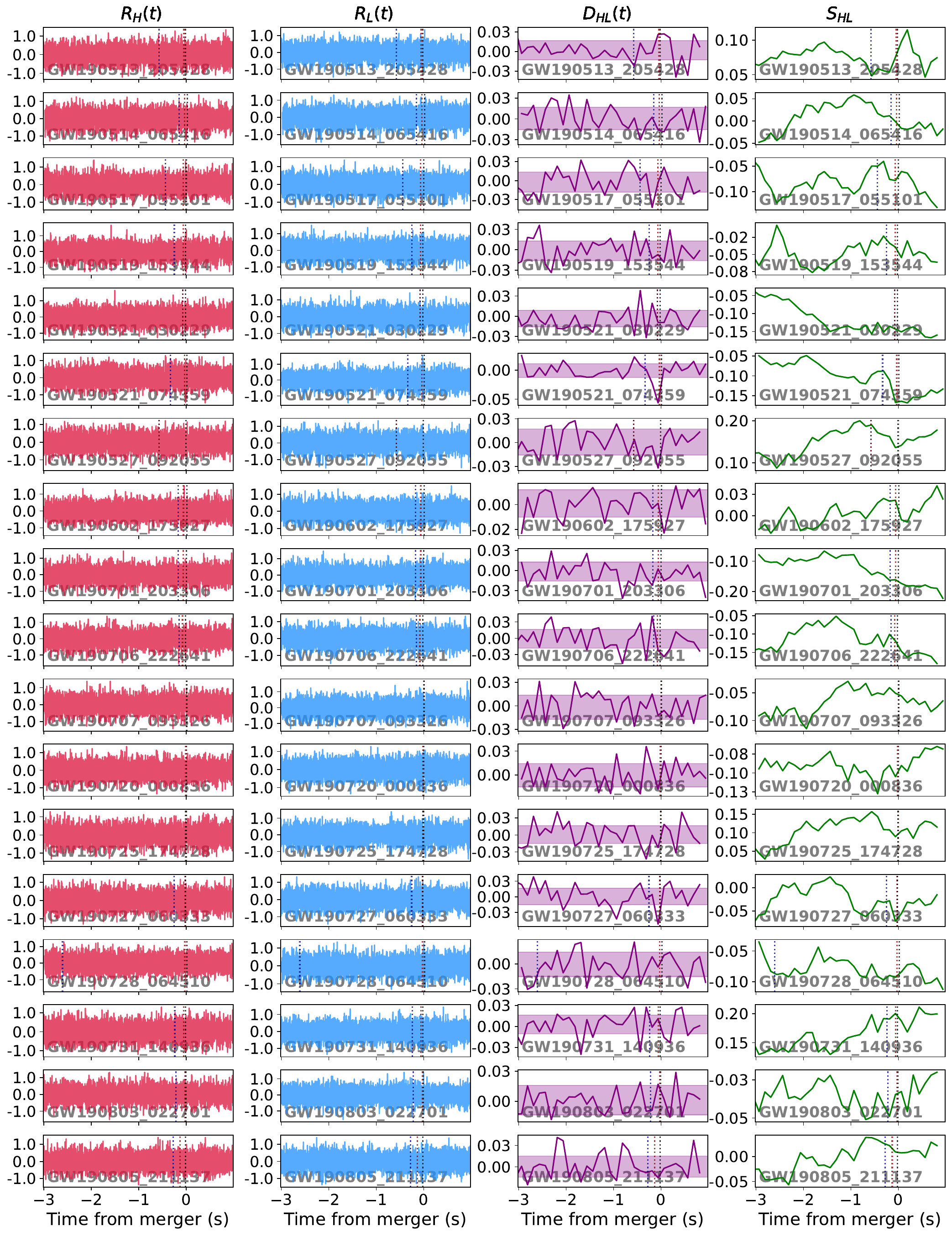}
    \caption{Performing $\mu$-\texttt{GLANCE} on the GW events dated between from May 13th, 2019 to August 5th, 2019. For each row, the panels represent the residual from H1, the residual from L1, the cross-correlation of these two residuals and the cumulative changes of the cross-correlation.}
    \label{fig:f2}
\end{figure}

\begin{figure}
    \centering
    \includegraphics[width=0.9\linewidth]{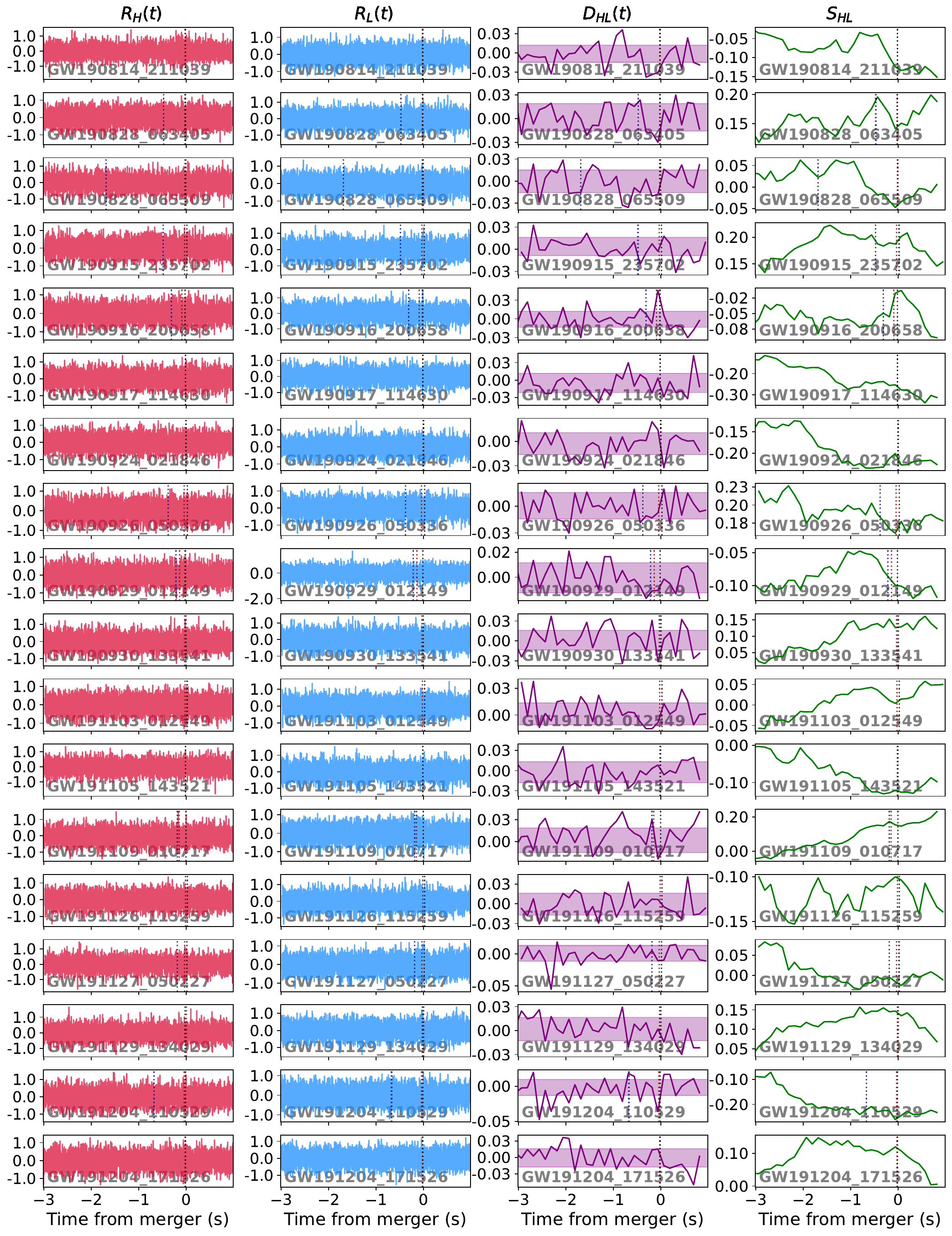}
    \caption{Performing $\mu$-\texttt{GLANCE} on the GW events dated between August 14th, 2019 to  December 4th, 2019. For each row, the panels represent the residual from H1, the residual from L1, the cross-correlation of these two residuals and the cumulative changes of the cross-correlation.}
    \label{fig:f3}
\end{figure}

\begin{figure}
    \centering
    \includegraphics[width=0.9\linewidth]{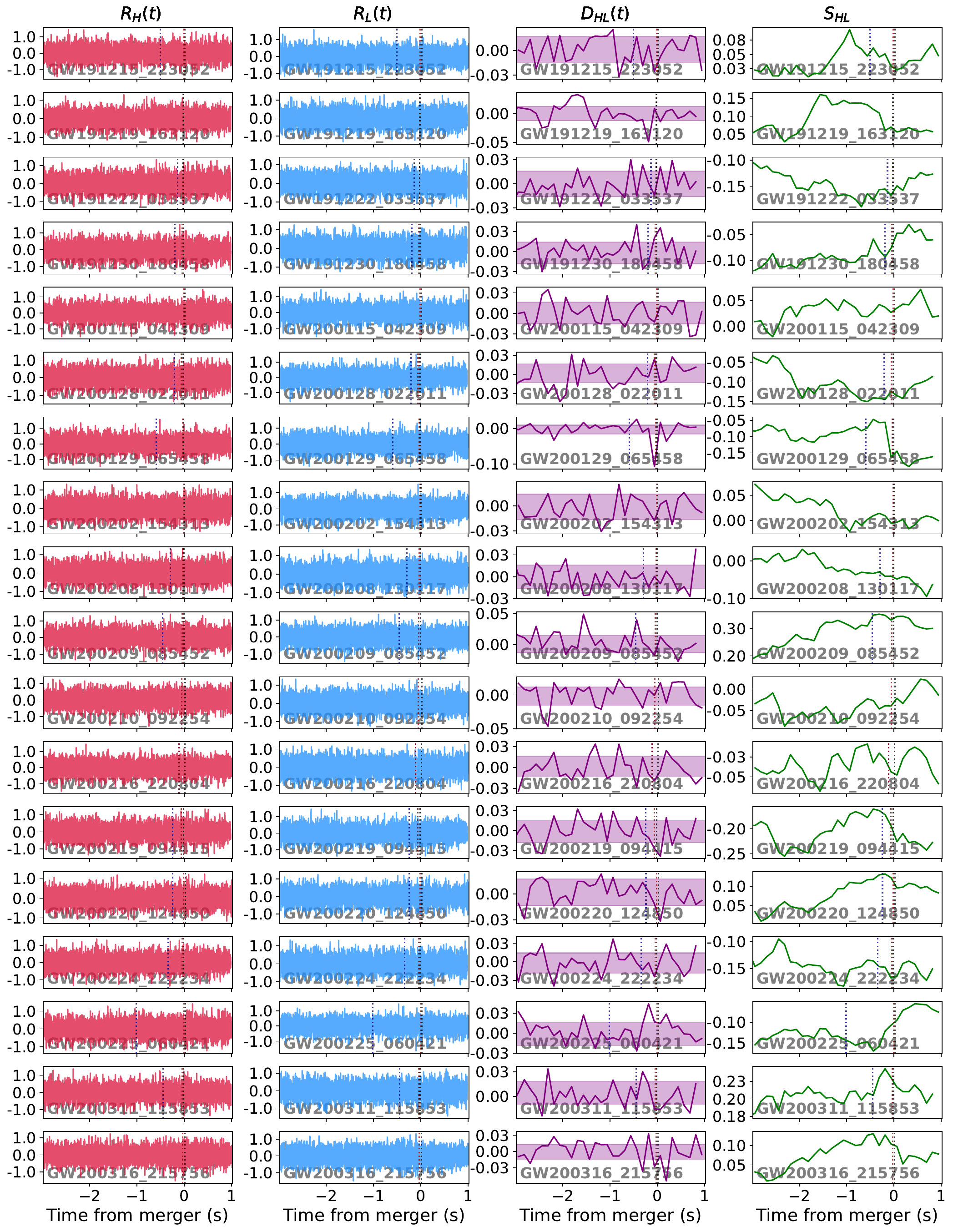}
    \caption{Performing $\mu$-\texttt{GLANCE} on the GW events dated between December 15th, 2019 to March 16th, 2020. For each row, the panels represent the residual from H1, the residual from L1, the cross-correlation of these two residuals and the cumulative changes of the cross-correlation.}
    \label{fig:f4}
\end{figure}

In this appendix, we present the performance of the  technique $\mu$-\texttt{GLANCE} for the events observed by both H1 and L1. 
In the figures \ref{fig:f1}, \ref{fig:f2}, \ref{fig:f3}, \ref{fig:f4} we present the analysis results on the complete 72 events whose data are publicly available. The events are arranged in a chronological order. The panels show the post-processed residual in the H1, post-processed residual in the L1, the cross-correlations of the two residuals and their localized nature and  build-up of the cross-correlation over time interval [$t_{20}$, $t_{\rm ISCO}$]  in the panels from left to right respectively. The purple horizontal band shows the noise cross-correlation based error. Through the stringent selection criteria mentioned in the subsection \ref{subsec:detect}, only one event $\rm GW190408\_1818102$ passes. One event in particular, $GW190521\_074359$ has a very high residual S/N of (-)4.41, it was rejected on the basis on the localized residual cross-correlation. On a similar basis, the event $\rm GW190413\_134308$ with a residual S/N of (-)2.96 rejected. One event $\rm GW200129\_065458$ having non-localized residuals having residual S/N close but below the threshold residual S/N: $\rho_{\rm residual} =  2.82 < \rho_{\rm residual } ^{\rm th}$ and thus is rejected. On a similar basis, the event $\rm GW190917\_114630$ is also rejected with a residual S/N of (-)2.57. These are a few events that were close to be chosen as interesting events for further checks but were not finally qualified. 

\end{document}